\documentclass{cernrep}
\usepackage{amssymb}
\usepackage{amsmath,epsfig}
\usepackage[latin1]{inputenc}
\usepackage{graphicx}
\usepackage[english]{babel}

\usepackage{xspace}
\usepackage{subfigure}
\usepackage{float}
\usepackage{multirow}
\usepackage{hyperref}
\usepackage{placeins}
\usepackage[percent]{overpic}

\title{RF Measurements of the New TOTEM Roman Pot}
   \author{O. Berrig, N. Biancacci, F. Caspers, A. Danisi,\\ J. Eberhardt, J. Kuczerowski, N. Minafra\footnote{nicola.minafra@cern.ch}, B. Salvant, C. Vollinger}

\begin{document}
\begin{table}
\vspace*{-1cm}
\mbox{
 \epsfig{file=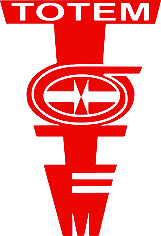,height=2cm} 
 \parbox[b]{13.1cm}{
   \begin{flushright}\textbf{	CERN-TOTEM-NOTE-2015-002\\ %
                            	August 2015}
   \end{flushright}
 }
}
\vspace*{0.5cm}
\end{table}

\maketitle

\begin{abstract}
The TOTEM experiment has been designed to measure the total proton-proton cross section and to study the elastic and diffractive scattering at the LHC energy. The measurement requires detecting protons at distances as small as 1 mm from the beam center: TOTEM uses Roman Pots, movable beam pipe insertions, hosting silicon detectors. In the first period of LHC operation no relevant problems were detected with Roman Pots retracted or inserted during special runs. However, when operating the LHC with high intensity beams, impedance induced heating has been observed during the Roman Pots insertion. In order to be compatible with the higher LHC beam current foreseen after the LS1, a new version of the Roman Pot has been proposed and optimized with respect to the beam coupling impedance. 
In this work we present the bench impedance measurements carried out on the new Roman Pot prototype. Single and double wire measurements, as well as probe measurements, were performed in order to detect possible harmful resonant modes. The laboratory setup has been as well simulated with the help of CST Particle Studio in order to benchmark the measurement results. Measurements and simulations are in close agreement confirming the equipment compatibility with the LHC requirements for safe operation.

\end{abstract}

\section{Introduction}
In previous works \cite{minafra2013rf} a new Roman Pot (RP) design was proposed and studied to cope with the increasing luminosity of the LHC after the LS1, together with an RF shield that can lower the impedance of the present RP avoiding the production of a completely new design. The main goal of the optimization was to reduce the beam induced heat that otherwise could severely damage the detectors inside the~RP.

This work describes the measurements that were done to benchmark the simulations and to understand their trustworthiness.
The only way to properly test the behaviour of the optimized RP is with the LHC beam. However, it is useful to test the model in the laboratory, using some techniques to reproduce the working conditions of the tested equipment.
Moreover, these tests allow to study the RF behaviour of the real equipment, without the need of numerical simulations, i.e. to understand the impact of the ferrite.
These tests are described in section \ref{testDescription} while sections \ref{Cylindrical} and \ref{Shielded} illustrate the results obtained for the two proposed designs.

\section{Description of the tests}
\label{testDescription}
In this section we describe the RF bench measurements that were setup in order to characterize the RP from an electromagnetic point of view.
The RF bench measurements on the device under test (DUT) were performed with single and double wires and with probes in operational working conditions. The presence of trapped modes, that may represent potentially harmful source of heating and beam instability in the LHC,  was carefully studied and bench-marked with numerical codes like CST Particle Studio~\cite{CSTcite} simulating the measurement setup.
Some of these tests were performed with the newly designed RP, together with several mechanical tests, before the installation of the final design in the LHC tunnel.

\subsection{Single wire test}
The single wire measurements are based on the idea that the field excitation induced by a current pulse travelling on a wire stretched along a beam pipe, is similar to the one induced by a traveling bunch of charged particles~\cite{Sands_1974}.  Inserting a thin wire in a vacuum chamber, we transform it in a coaxial transmission line. The transmission scattering coefficient $S_{21}$ of the line can be therefore measured using a vectorial network analyzer (VNA) and provides an observable for those fields having an electrical  longitudinal component at the centre of the beam pipe.  This method is commonly adopted to measure the longitudinal beam coupling impedance of a device.
\par
\begin{figure}[htb!]
\centering
\includegraphics[width=0.9\textwidth]{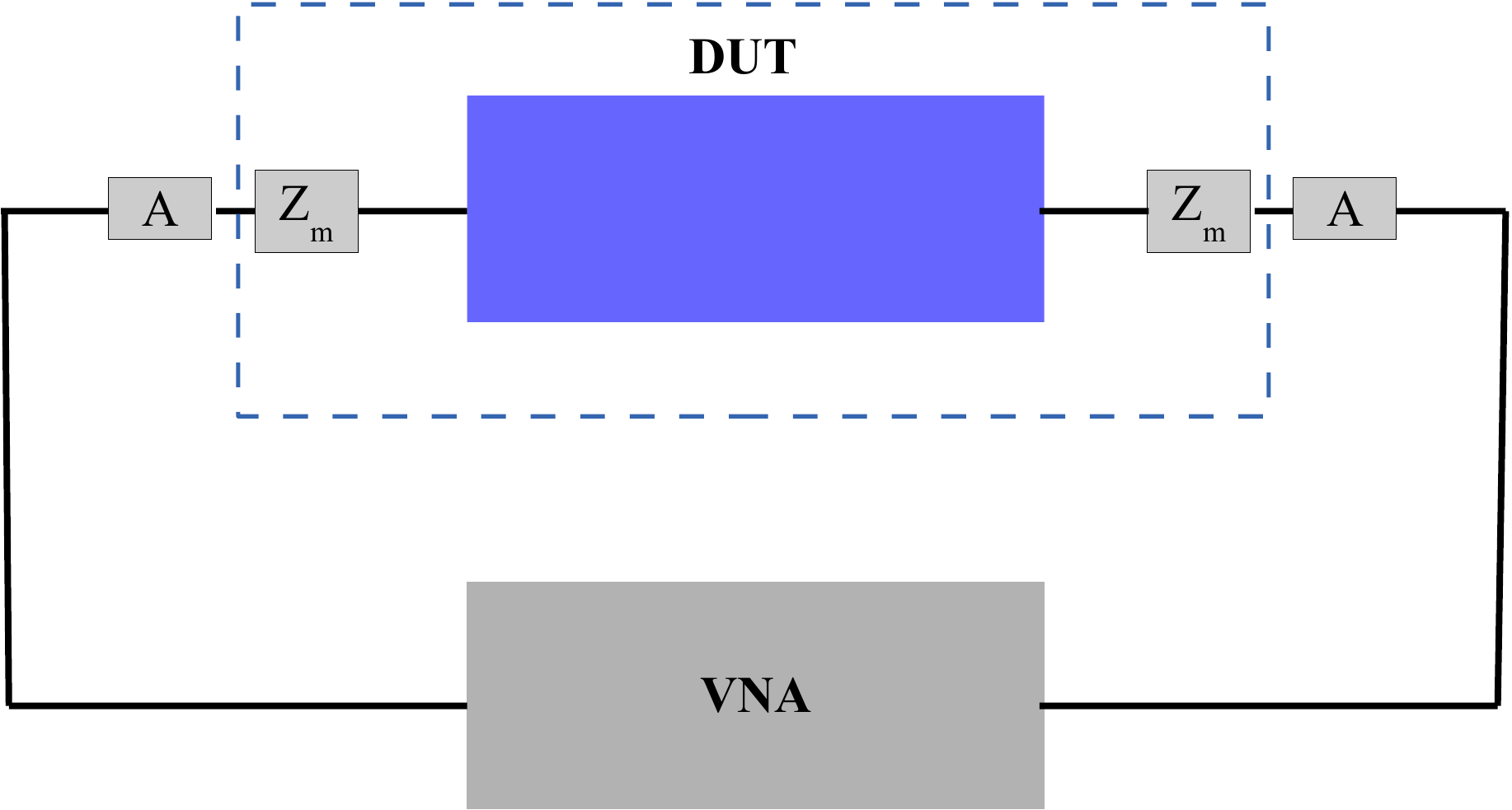}
\caption{Single wire setup: a single wire is stretched along the device under test (DUT). The impedance matching is ensured by the network (block ``$Z_m$'') and the series attenuator (block ``A''). }
\label{wireTest_setup}
\end{figure}

 To perform the measurements, the coaxial line made by the wire and the beam pipe has to be adapted to the \Unit{50}{\Omega} impedance of the VNA cables. This can be done using a matching network. As a simple and practical solution, we considered a single resistor $Z_m$ in series to the characteristic impedance $Z_c$ of the coaxial line before and after the DUT, as shown in fig. \ref{wireTest_setup}. The reflections can be also reduced adding an attenuator in series to the matching resistor.

\par The coaxial line characteristic impedance $Z_c$ is given by the formula~\cite{collin2007foundations}
\begin{equation}
Z_{c}\simeq 60 \Omega \, \ln \frac{D}{d},
\end{equation}
where $D$ is the diameter of the vacuum pipe and $d$ is the diameter of the wire. In our setup, the wire diameter $d$ is \Unit{0.5}{mm} while the vacuum pipe diameter $D$ is \Unit{80}{mm}. The characteristic impedance is therefore $Z_{c} \simeq 304 \Omega$. To match the \Unit{50}{\Omega} of the network analyzer a $Z_m=Zc-50\Omega=$\Unit{254}{\Omega} resistor is needed. Several resistors were tested to find the ones with the closest value to \Unit{254}{\Omega}. However, the closest ones commercially available with 5\% tolerance has a resistance of \Unit{240}{\Omega}. 

\par
\begin{figure}[htb!]
\centering
\includegraphics[width=0.9\textwidth]{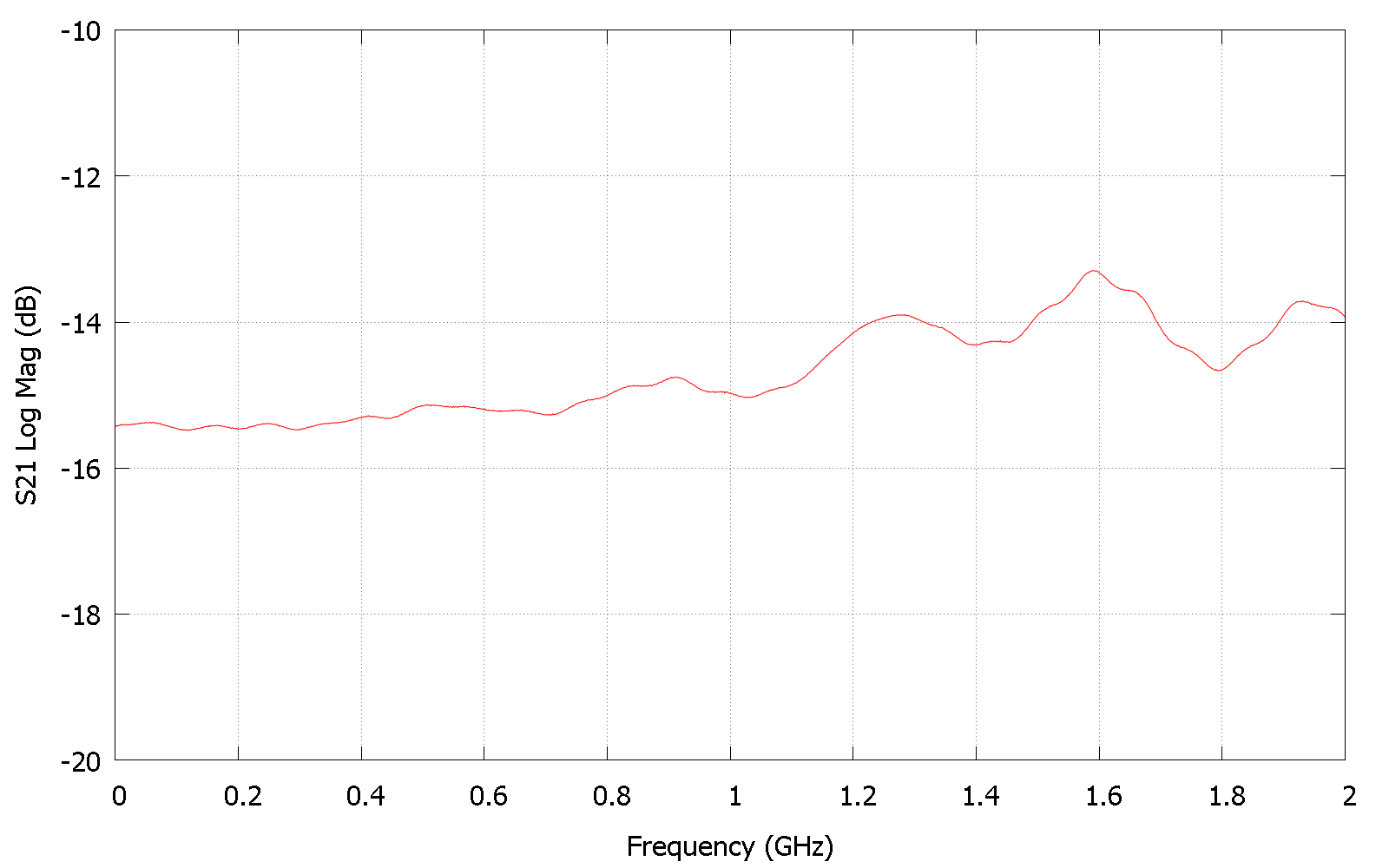}
\caption{$S_{21}$ for the wire test with a simple beam pipe.}
\label{BeamPipeWire}
\end{figure}

 A first test has been done using the single wire setup  to test a vacuum pipe of the same diameter as the RP. Looking at the $S_{21}$ in fig. \ref{BeamPipeWire}, it is possible to note some reflections due to a residual mismatch despite the careful selection of the resistors: from a practical point of view it is difficult to achieve a better matching in this measurement setup since systematic effects, like the additional losses introduced by the cables and the wire alignment with respect to the vacuum pipe center, are difficult to control. Since no attenuators were used in the setup, the attenuation level is $\simeq -8~$dB.

\par An example of measurement setup on a RP is shown in fig~\ref{BeamPipeWire}: the 10 dB attenuator are in series to the matching resistor connected to the wire and contained in the sucobox.

\begin{figure}[htb!]
\centering
\includegraphics[width=0.9\textwidth]{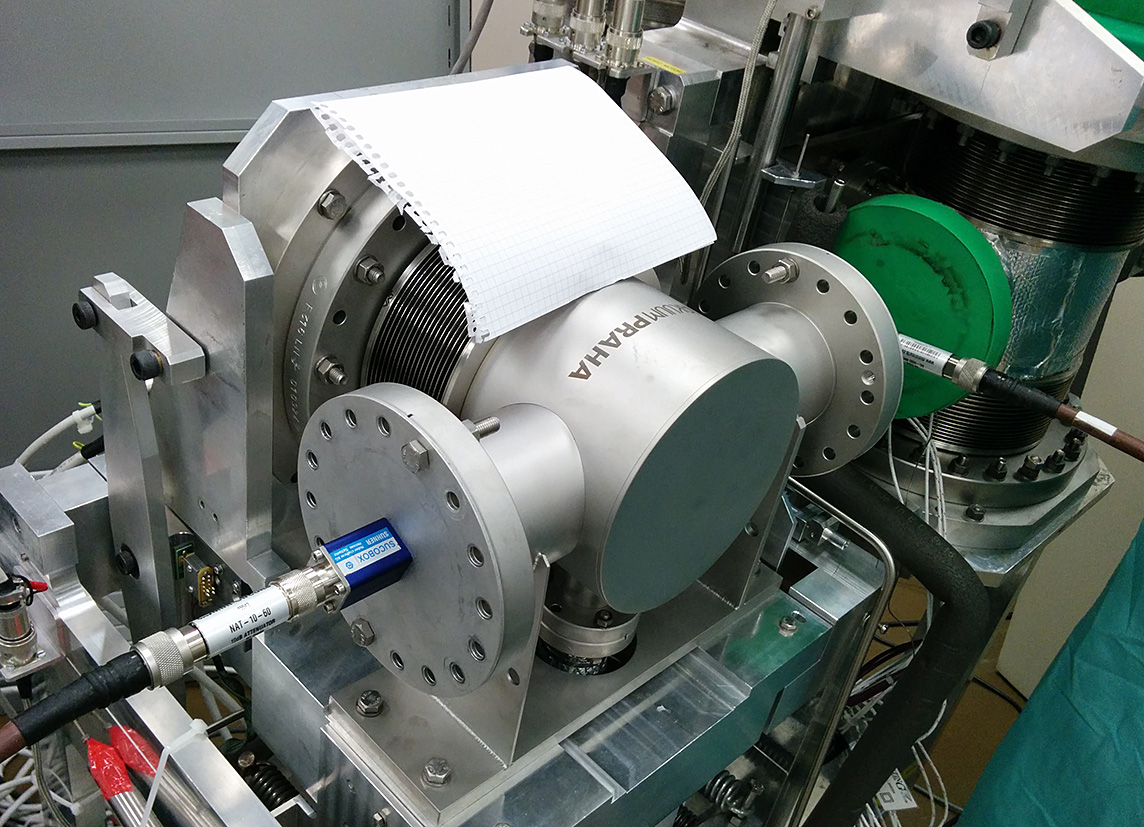}
\caption{Photo of the setup for the wire test using a prototype RP.}
\label{WireTest}
\end{figure}

\subsection{Double Wire test}
\label{DoubleWireDescription}
\begin{figure}[htb!]
\centering
\includegraphics[width=0.9\textwidth]{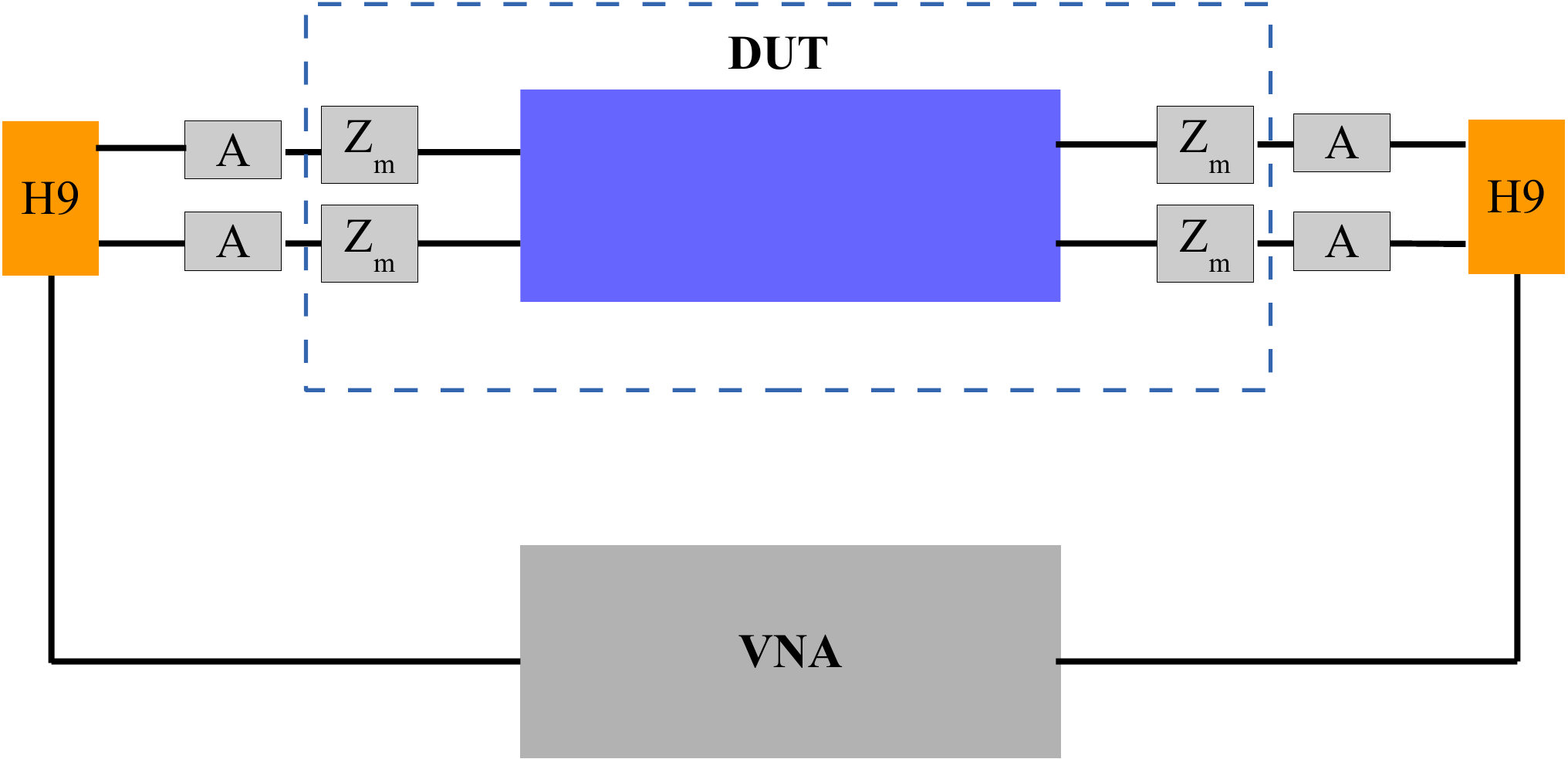}
\caption{Double wire setup: two wires are stretched along the device under test (DUT). A hybrid unit (block ``H'') splits and dephases the VNA signal by $\pi$. The impedance matching is ensured by the network (block ``Z'') and the series attenuator (block ``Att''). }
\label{Double_wire}
\end{figure}

To be able to see the resonances with a transverse component, it is possible to use two parallel wires as shown in fig.~\ref{Double_wire}. The current in the two wires is $\pi$ out of phase in order to ``simulate'' the beam dipolar moment~\cite{Nassibian_1977}.

Inserting two wires in the DUT, we make a two-wires shielded  transmission line. Having a three-conductors line, we can support two TEM modes, so called \emph{odd} when the wire currents are $180^\circ$ out of phase between each other, and \emph{even} mode, when the wire currents are in phase. Using an hybrid module, we can split the signal coming from the VNA and give a $180^\circ$ phase shift. We excite, therefore, only the odd mode, whose characteristic impedance is given by~\cite{h1992taschenbuch}:

\begin{equation}
Z^\prime_c=\frac{Z_0}{\pi}\left(\log\left(2 p \frac{1-q^2}{1+q^2}\right)-\frac{(1+4 p^2)(1-4 q^2)}{16 p^4}\right)
\end{equation}

with $Z_0\simeq 377\Omega$ the vacuum characteristic impedance, $s$ is the distance between the wires, and defining the parameters $p=s/d$ and $q=s/D$ . This impedance refers to the three-conductor (two wires and the beam pipe as shield) transmission line. 
\par In order to match the impedance to the two-line (single wire and beam pipe as shield) impedance of the VNA we need to calculate the characteristic impedance of each of the two stretched wire. Since we support an \emph{odd} mode a virtual ground plane can be placed between the wires without perturbing the field pattern. The characteristic impedance of the new wire-ground plane transmission line will be $Z_c=Z^\prime_c/2$.

As for the single wire case, we can now match each of the virtual two-conductor lines to the VNA. Considering a wire spacing of $s=1$~cm we have $Z_c\simeq 220\Omega$. The matching resistors have been therefore chosen to be as close as possible to $Z_m=Z_c-50\Omega=170\Omega$. 

The double wire measurement on the RP has been performed both with the two wires perpendicular to the bottom plane (horizontal) of the RP and with the two wires parallel to it.

\subsection{Probe test}
\begin{figure}[htb!]
\centering
\includegraphics[width=0.9\textwidth]{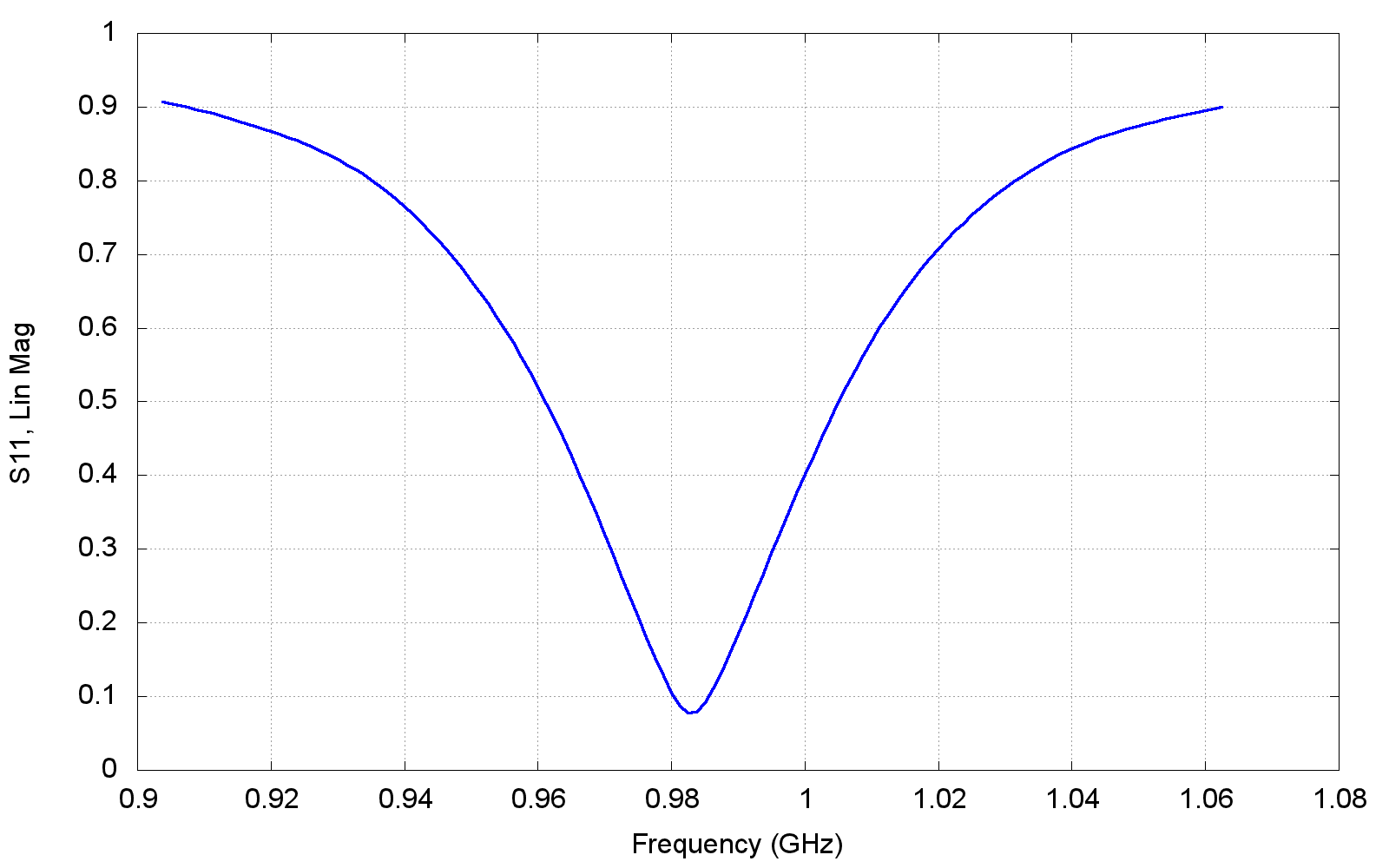}
\caption{Reflection $S_{11}$ scattering parameter around a resonance. To compute the merit factor we select $f_0$, the frequency corresponding to the minimum.}
\label{probef0}
\end{figure}

\begin{figure}[t!]
\centering
\begin{overpic}[width=0.45\textwidth]{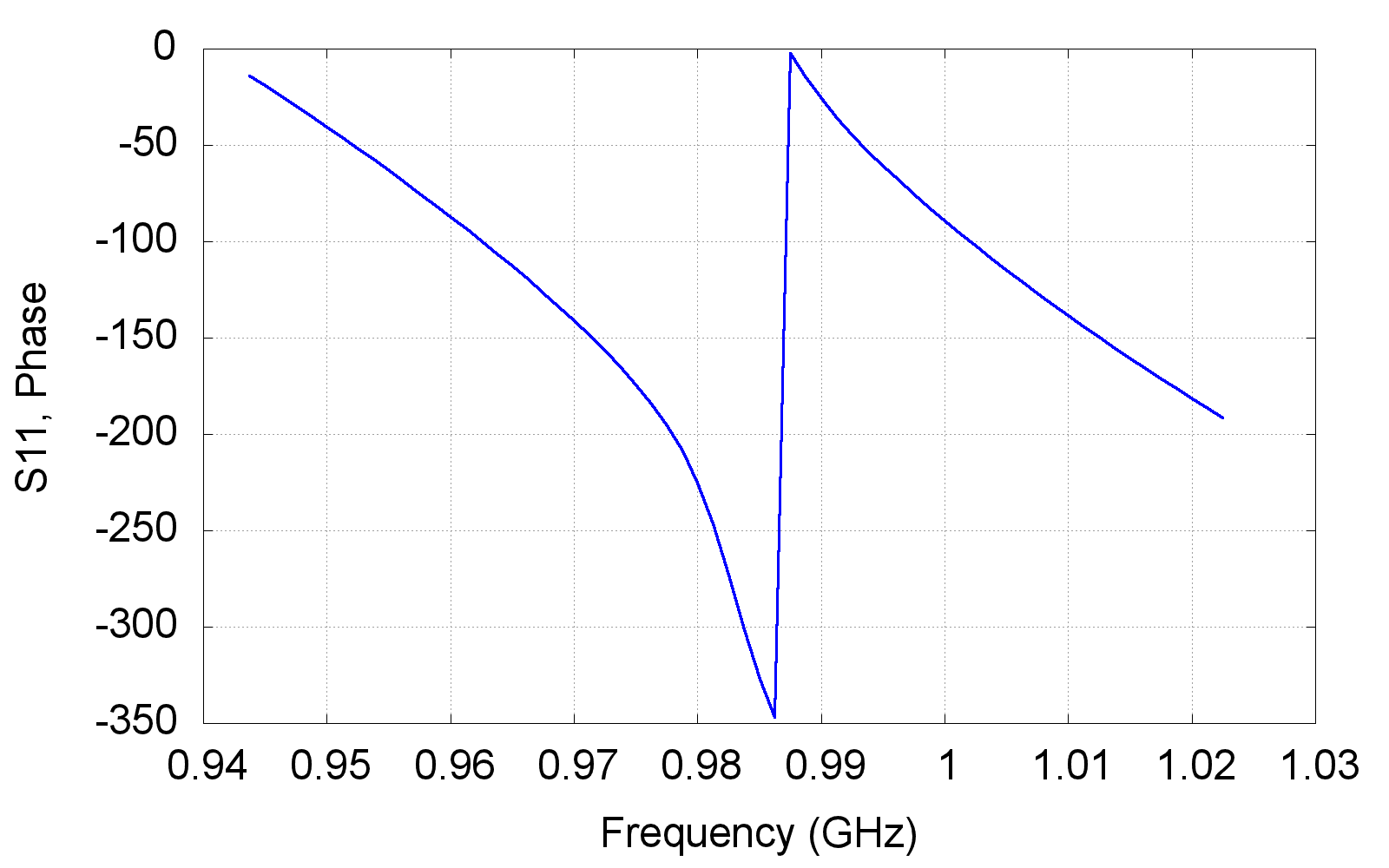}
\put (60,12) {\footnotesize wrong \emph{electrical delay}}
\end{overpic}
\quad
\includegraphics[width=0.45\textwidth]{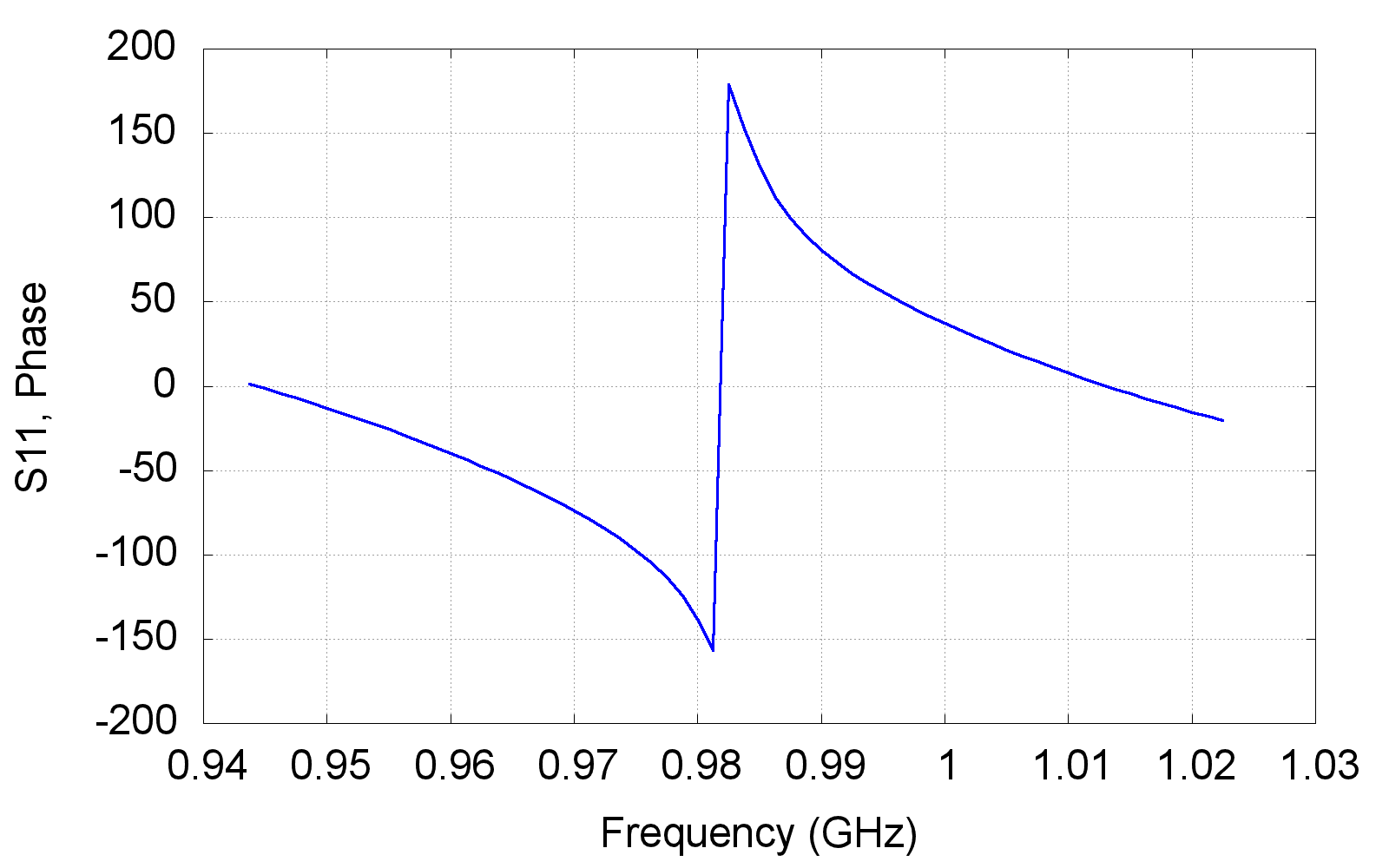}
\caption{Phase of $S_{11}$ around a resonance. The graph on the left shows an example of measurement with a wrong \emph{electrical delay}. The error can be corrected through the VNA to obtain the graph on the right.}
\label{phase_delay}
\end{figure}

\begin{figure}[htb!]
\centering
\vspace{-7cm}
\includegraphics[width=0.45\textwidth]{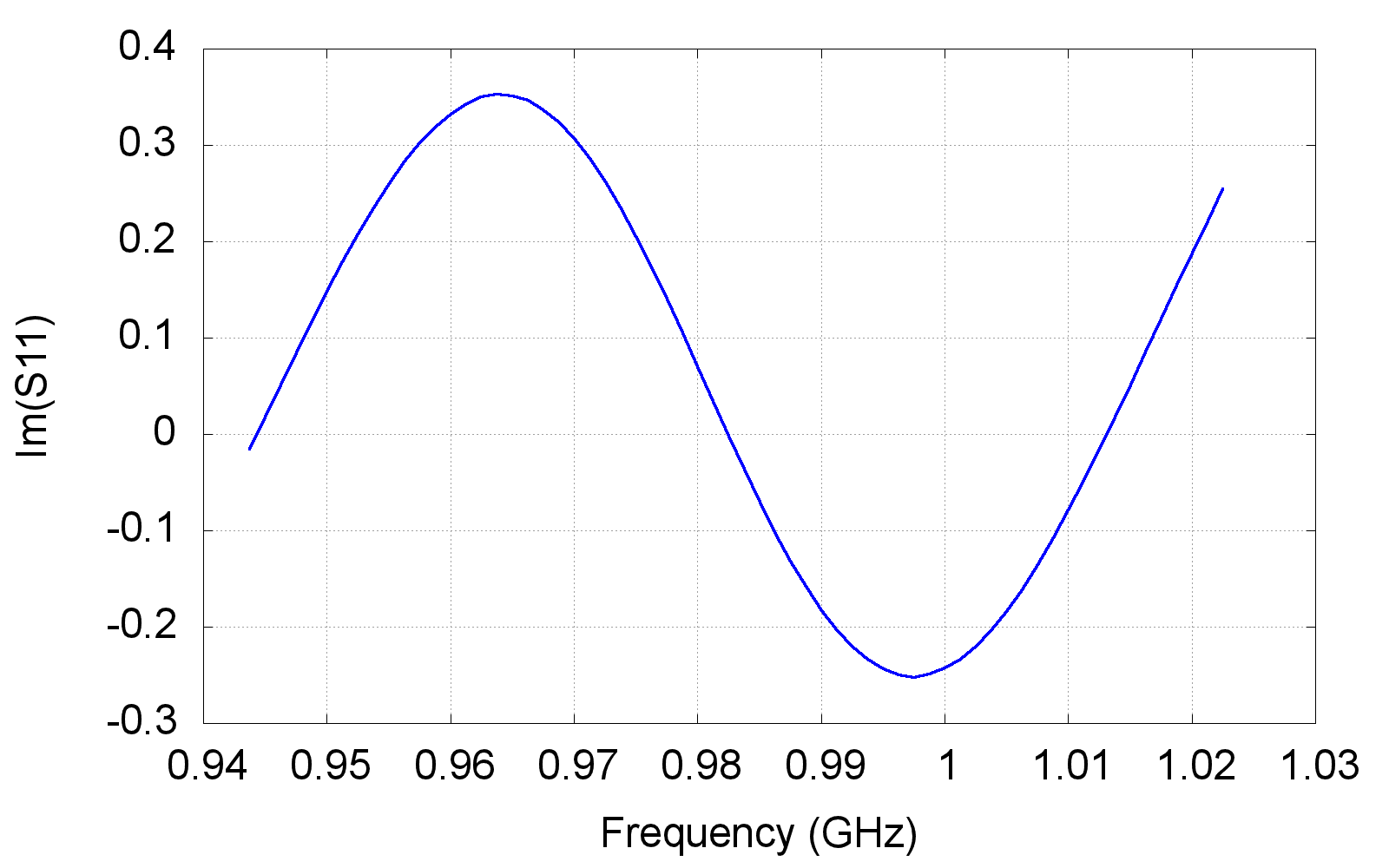}
\caption{Imaginary part of $S_{11}$ around a resonance. To compute the merit factor we select the frequencies $f_1$ and $f_2$ at the local maximum and minimum $Im(S_{11})$.}
\label{probef1f2}
\end{figure}

A different and complementary approach to the wire measurements, is using a probe to excite the DUT and measure the reflection coefficient $S_{11}$ from the access ports. Even if also the probe will excite self-resonances, these can be disentangled by the DUT ones estimating  the reflection coefficient for several positions: if the detected resonances will not move, these can be associated to the DUT, otherwise they are linked to the probes.
During the measurements, we moved slowly the probe inside the RP through the beam pipe aperture until a resonance could be localized. If the resonance is associated with the RP, the frequency and merit factor are therefore calculated.\\
The procedure used to compute the unloaded $Q_0$ from the $S_{11}$ using the VNA is the {following~\cite{Caspers_2007,Eberhardt_2014}:}
\begin{enumerate}
\item pick the resonance $S_{11}$ central frequency $f_0$ (fig. \ref{probef0}),
\item add an \emph{electrical delay} and/or a \emph{phase offset} to flatten the phase of the $S_{11}$ (fig. \ref{phase_delay}),
\item pick the local maximum $f_1$ and minimum $f_2$ of the $Im(S_{11})$ (fig. \ref{probef1f2}),
\item estimate the probe coupling $\beta = \dfrac{1 - |S_{11}(f_0)|}{1 + |S_{11}(f_0)|}$, if $\beta \leq 1$ or $\beta = \dfrac{1 + |S_{11}(f_0)|}{1 - |S_{11}(f_0)|}$, if $\beta \geq 1$
\item compute the loaded merit factor $Q_L = \dfrac{f_0}{f_2 - f_1}$,
\item compute the unloaded merit factor $Q_0 = Q_L (1 + \beta)$.
\end{enumerate}

\clearpage
\section{Cylindrical RP}
\label{Cylindrical}
\begin{figure}[htb!]
\centering
\includegraphics[width=0.45\textwidth]{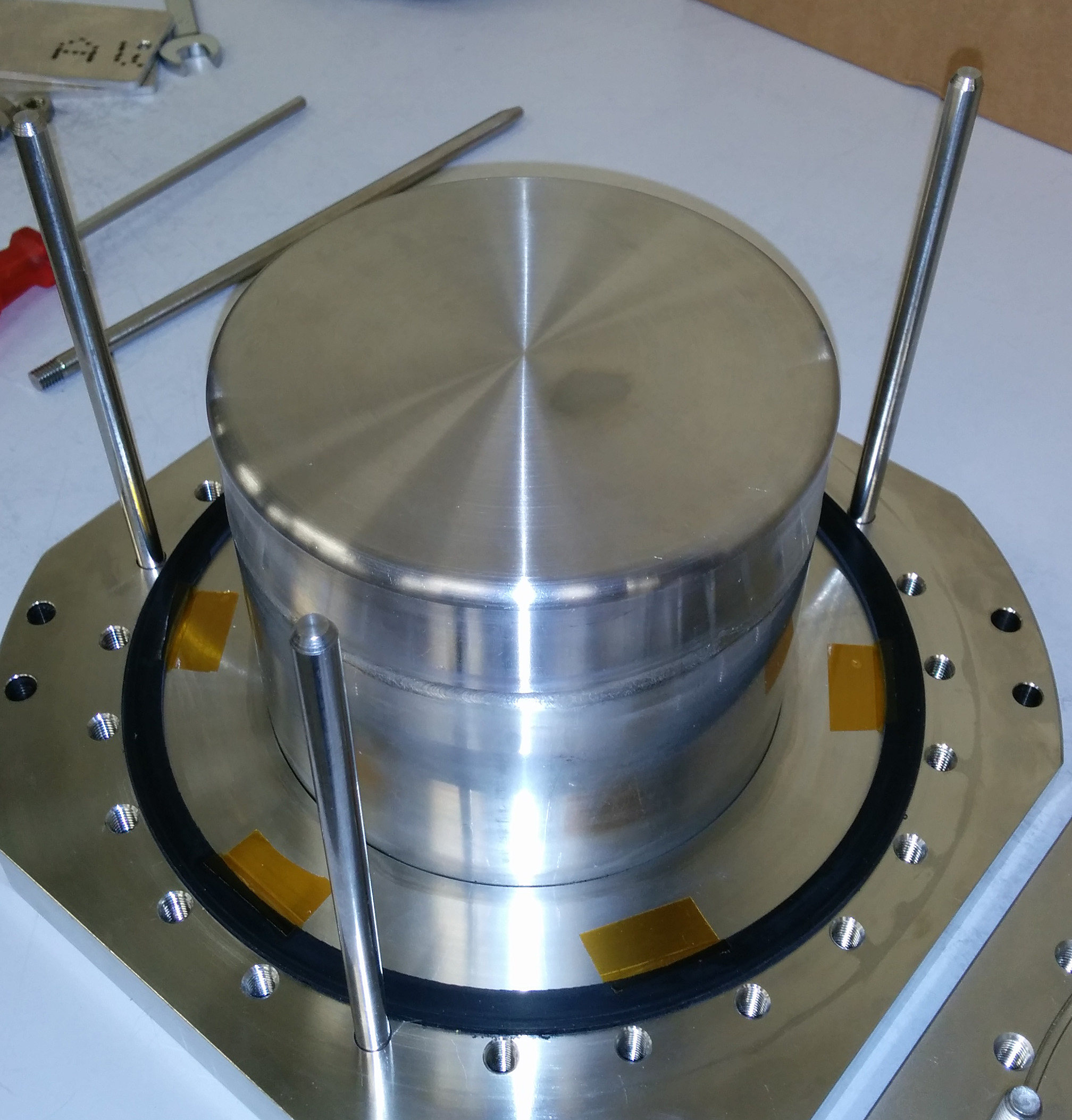}\quad
\includegraphics[width=0.45\textwidth]{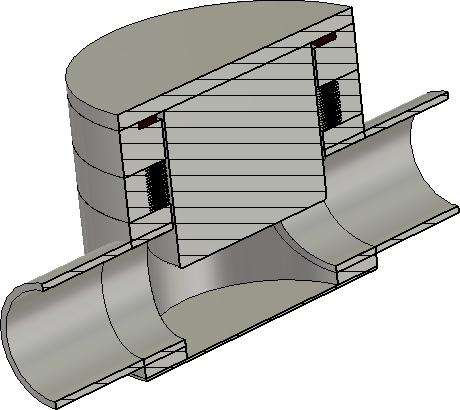}
\caption{The prototype used for the measurements and the model used for the simulations of the cylindrical RP.}
\label{CylindricalRP}
\end{figure}

The cylindrical RP has been designed to optimize the impedance of the Totem RP and, at the same time, increase the space available for the detector package inside the RP.
The prototype was realized to make mechanical and RF tests and it is shown in fig. \ref{CylindricalRP}.

\subsection{Single wire test}
The wire test has been done for various position of the RP but the most relevant are the ones with the RP in garage position (40 mm from the centre) and at 10 mm from the centre of the beam pipe.
With the RP closer to the wire the interference with the measurement device becomes too big; moreover, with the probe and the double wire, it was physically not possible to insert the RP so close because of the size of the probe and the mechanism to mount the two parallel wires; hence, to have a complete picture of the system we ran all the tests at 10 mm.

\begin{figure}[htb!]
\centering
\begin{overpic}[width=0.8\textwidth]{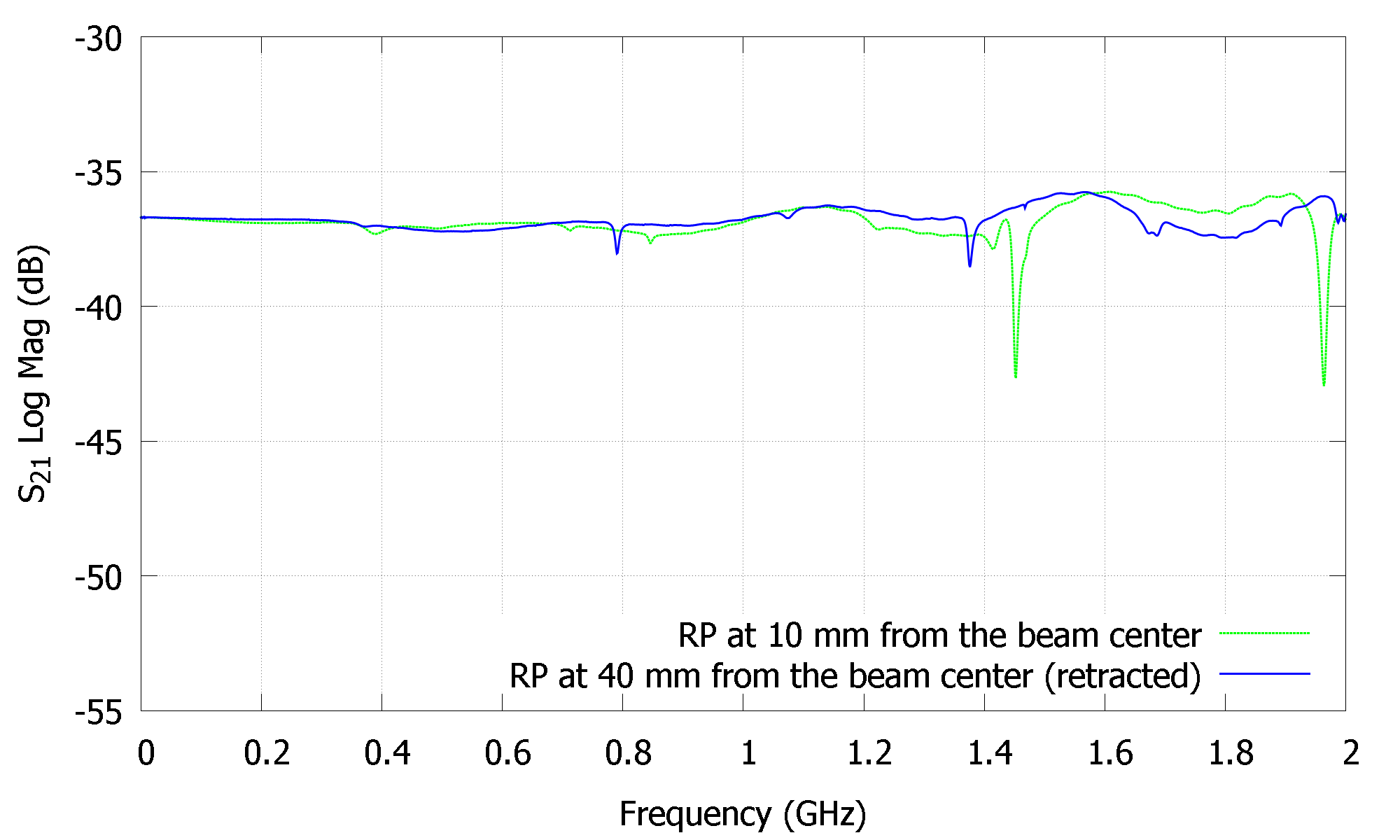}
\put (12,10) {\footnotesize Single wire measurements}
\end{overpic}
\caption{$S_{21}$ for the wire test with the Cylindrical RP without ferrite at 10 mm and 40 mm from the centre of the beam pipe.}
\label{CylindricalWOFerrite}
\end{figure}

\begin{figure}[htb!]
\centering
\begin{overpic}[width=0.8\textwidth]{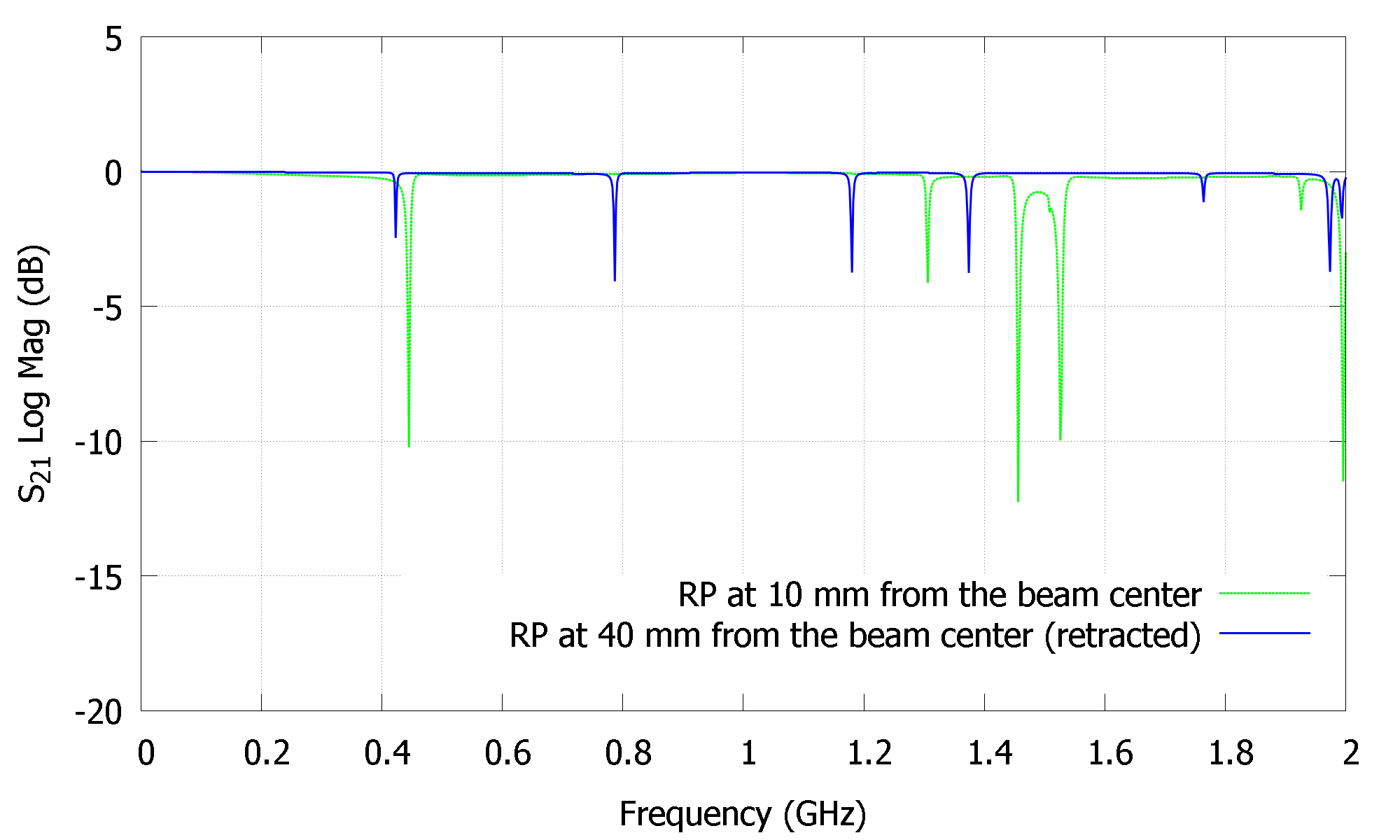}
\put (12,10) {\footnotesize Single wire simulations}
\end{overpic}
\caption{Simulated $S_{21}$ for the wire test with the Cylindrical RP without ferrite at 10 mm and 40 mm from the centre of the beam pipe.}
\label{CylindricalWOFerrite_SIM}
\end{figure}

Fig. \ref{CylindricalWOFerrite} and fig. \ref{CylindricalWOFerrite_SIM} show $S_{21}$ without using the ferrite.
The measurements done without ferrite are useful to check the reliability of the simulations: the very high Q of the resonances make them easily detectable.
Indeed, comparing the two plots it is possible to note that almost all resonances are present both in measurements and simulations and their frequency is compatible.
The peaks in the simulations are sharper because all losses, i.e. the resistance of the wire, are ignored.

\clearpage

\begin{figure}[htb!]
\centering
\begin{overpic}[width=0.8\textwidth]{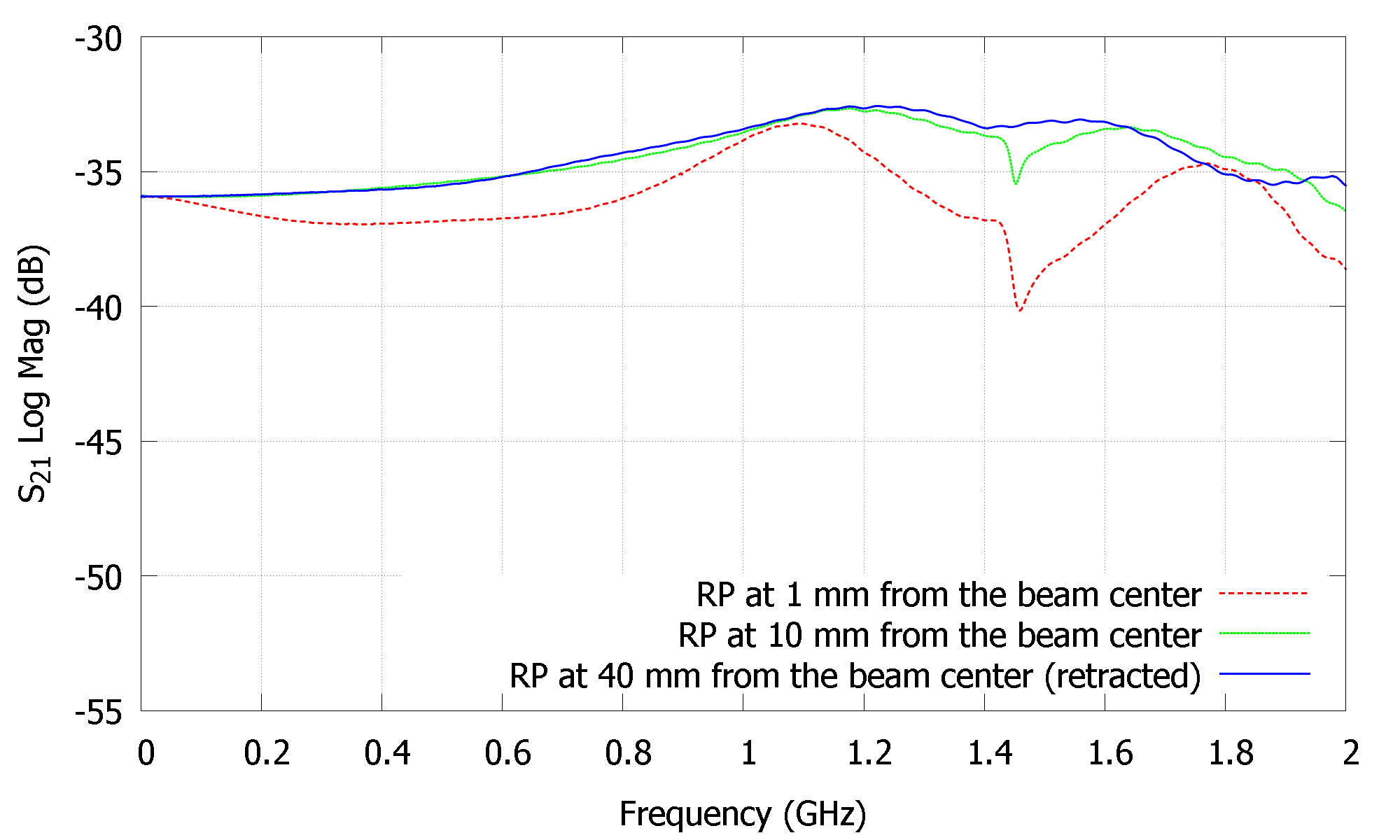}
\put (12,60) {\footnotesize Single wire measurements}
\end{overpic}
\caption{$S_{21}$ for the wire test with the Cylindrical RP with ferrite at 1~mm, 10~mm and 40~mm from the centre of the beam pipe.}
\label{CylindricalWFerrite}
\end{figure}

\begin{figure}[htb!]
\centering
\begin{overpic}[width=0.8\textwidth]{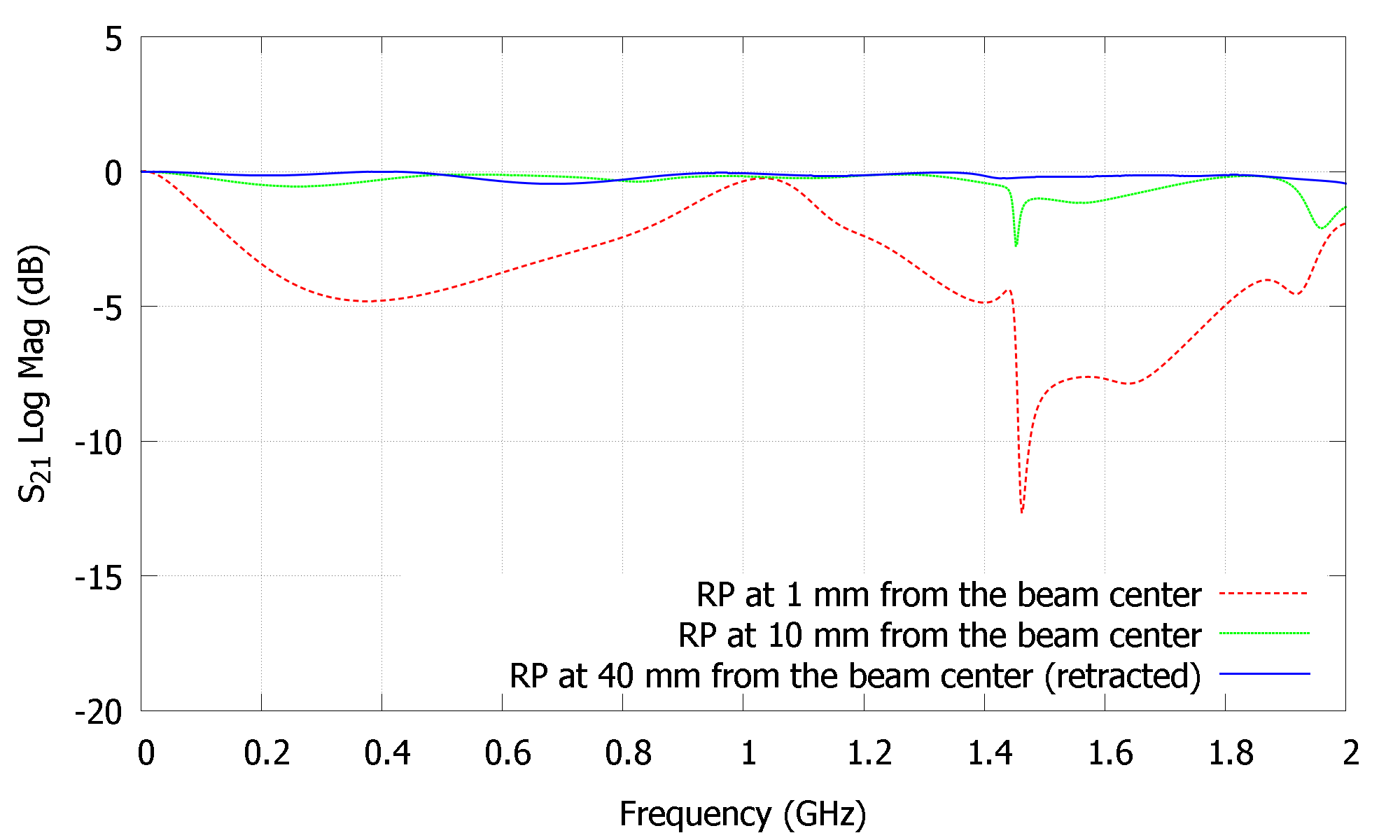}
\put (12,60) {\footnotesize Single wire simulations}
\end{overpic}
\caption{Simulated $S_{21}$ for the wire test with the Cylindrical RP with ferrite at 1~mm, 10~mm and 40~mm from the centre of the beam pipe.}
\label{CylindricalWFerrite_SIM}
\end{figure}

\clearpage
Fig. \ref{CylindricalWFerrite} and fig. \ref{CylindricalWFerrite_SIM} show $S_{21}$ after the mounting of the ferrite ring.
The effectiveness of the ferrite is clear: there are no more detectable peaks below 1.4~GHz. Also in this case, the agreement with the simulation can be considered acceptable. 

\clearpage
\subsection{Double wire test}
\begin{figure}[htb!]
\centering
\begin{overpic}[width=0.8\textwidth]{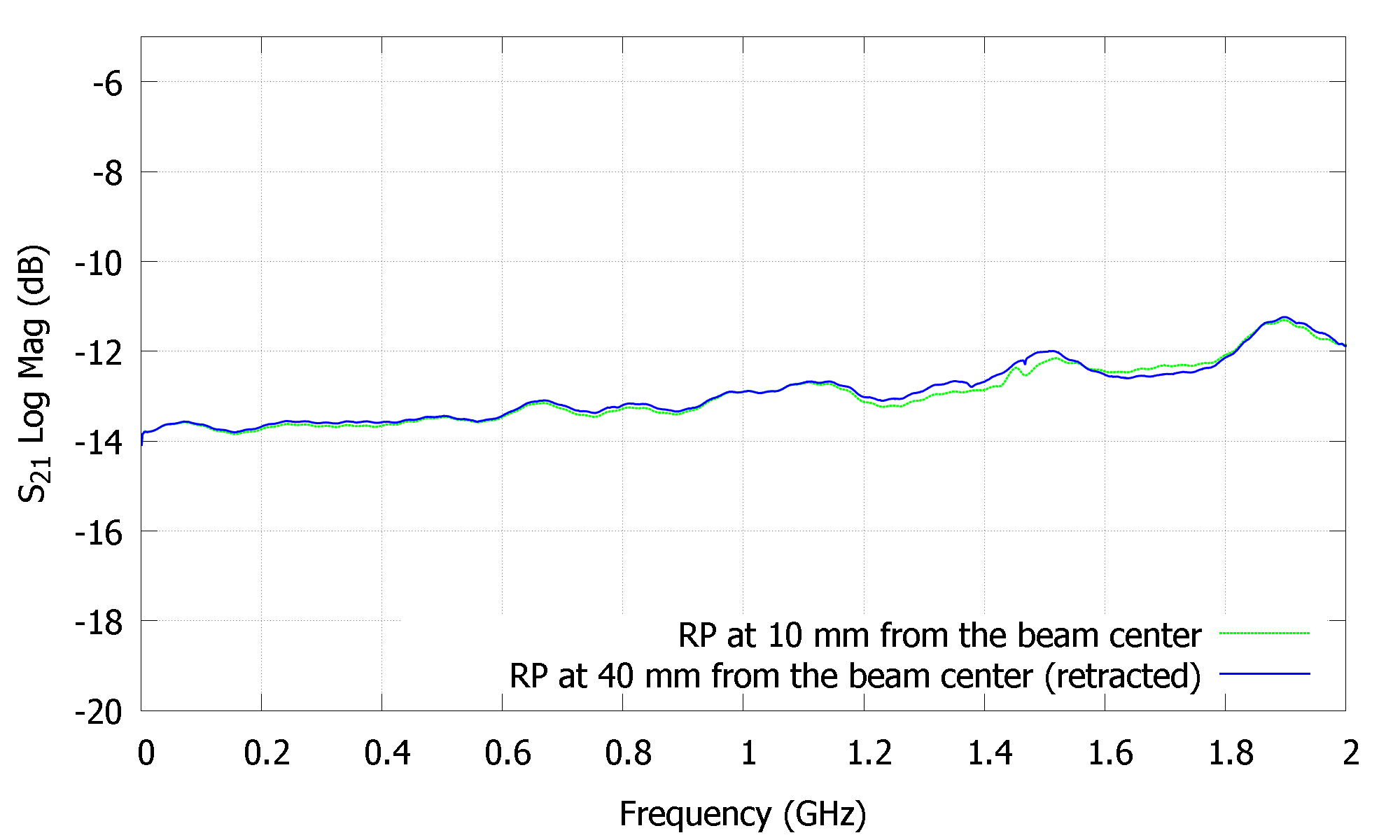}
\put (12,60) {\footnotesize Double wire horizontal}
\end{overpic}
\caption{$S_{21}$ for the test with two wires horizontal with the Cylindrical RP without ferrite at 10 mm and 40 mm from the centre of the beam pipe.}
\label{CylindricalWOFerrite_HOR}
\end{figure}

\begin{figure}[htb!]
\centering
\begin{overpic}[width=0.8\textwidth]{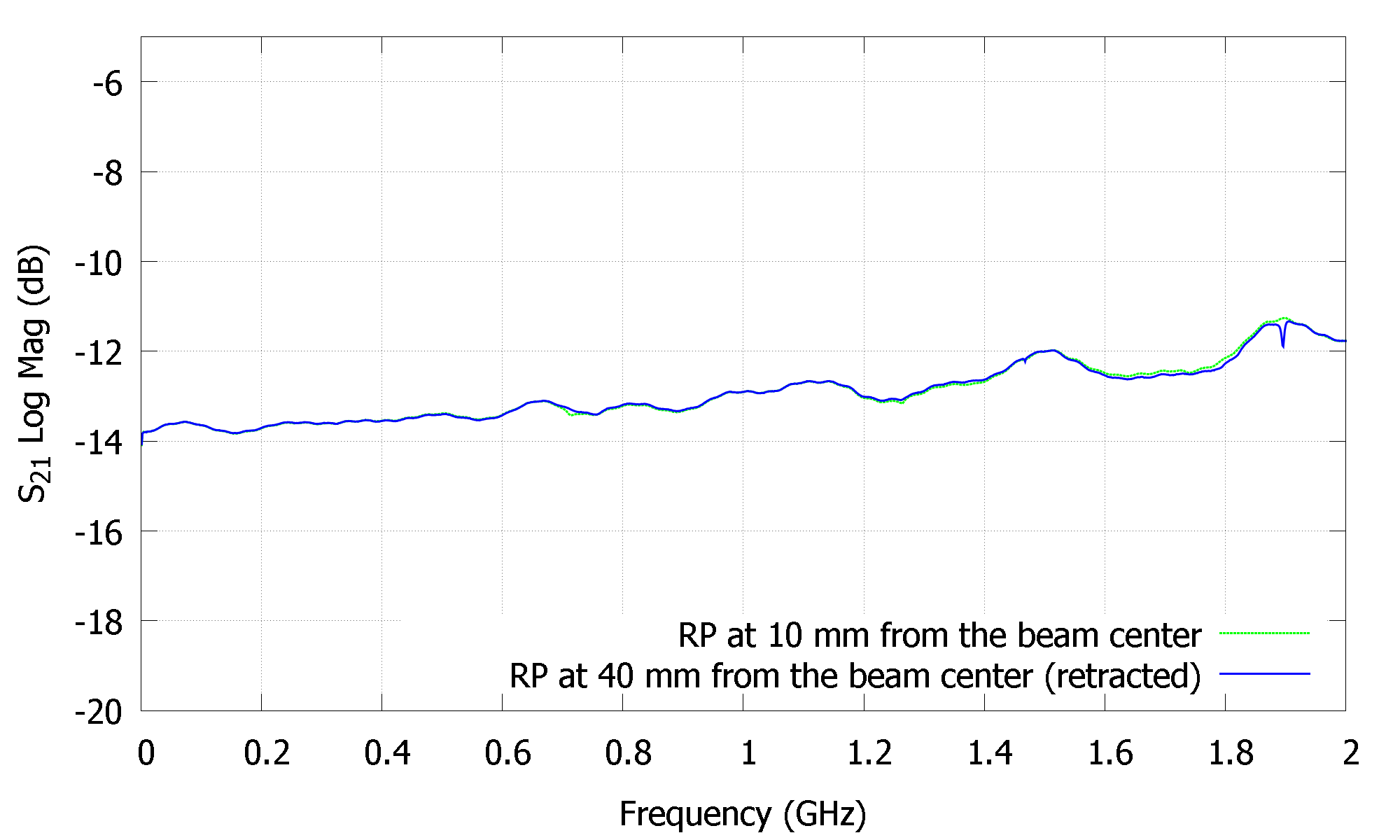}
\put (12,60) {\footnotesize Double wire vertical}
\end{overpic}
\caption{$S_{21}$ for the test with two wires vertical with the Cylindrical RP without ferrite at 10 mm and 40 mm from the centre of the beam pipe.}
\label{CylindricalWOFerrite_VER}
\end{figure}

\begin{figure}[htb!]
\centering
\begin{overpic}[width=0.8\textwidth]{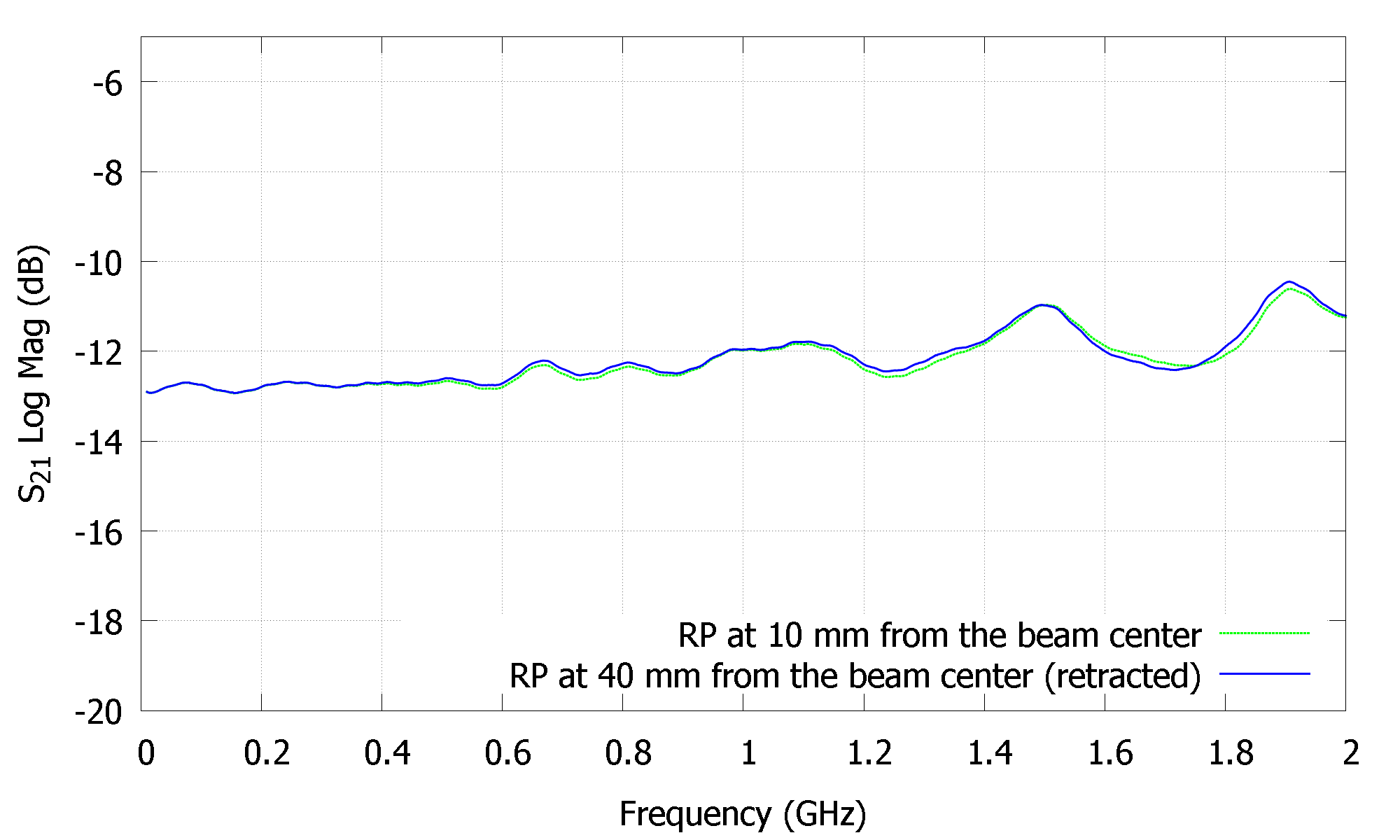}
\put (12,60) {\footnotesize Double wire horizontal, with ferrite}
\end{overpic}
\caption{$S_{21}$ for the test with two wires horizontal with the Cylindrical RP with ferrite at 10 mm and 40 mm from the centre of the beam pipe.}
\label{CylindricalWFerrite_HOR}
\end{figure}

\begin{figure}[htb!]
\centering
\begin{overpic}[width=0.8\textwidth]{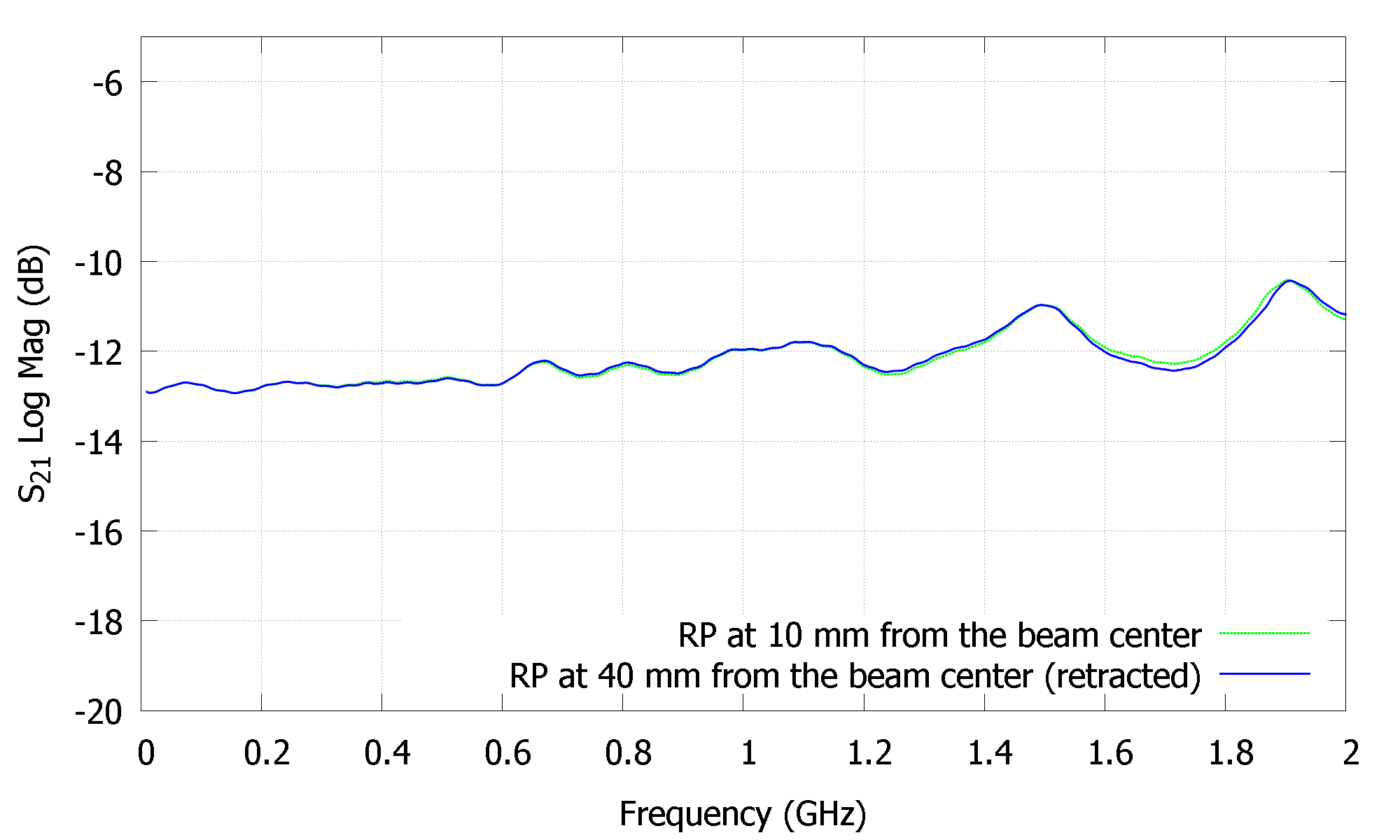}
\put (12,60) {\footnotesize Double wire vertical, with ferrite}
\end{overpic}
\caption{$S_{21}$ for the test with two wires vertical with the Cylindrical RP with ferrite at 10 mm and 40 mm from the centre of the beam pipe.}
\label{CylindricalWFerrite_VER}
\end{figure}

Fig. \ref{CylindricalWOFerrite_HOR} and fig. \ref{CylindricalWOFerrite_VER} show $S_{21}$ measured as described in section \ref{DoubleWireDescription}, using the cylindrical RP without ferrite. Both in the vertical and horizontal plane only a small resonance is visible at $\sim$~1.4~GHz.
This resonance is effectively damped by the ferrite, as shown in fig. \ref{CylindricalWFerrite_HOR} and fig. \ref{CylindricalWFerrite_VER}.
No other relevant transverse modes have been detected with this method.

\clearpage
\subsection{Probe test}
The probe test was done with the Cylindrical RP without ferrite in garage position. After the installation of the ferrite, in garage position it was not possible to detect any resonance; therefore, with the ferrite, only results with the RP inserted at 10 mm from centre are shown.
All the resonances measured in the tests were predicted by the simulation. Some resonances are not detected using the probe because some of them have a null electrical field in the centre of the beam-pipe, where the probe is positioned.

Table \ref{TableCyl40WOFerrite} and table \ref{TableCyl10WFerrite} show a good agreement between the measurements and the simulations regarding the frequency. The discrepancy between the measured and simulated Q is due to the losses in connectors, adapters and cables used in the real setup and not considered in the simulations.

\begin{table}[htb!]
\centering
\begin{tabular}{ | c c c | c c | }
  \hline        
  \multicolumn{3}{|c|}{Probe measurement} & \multicolumn{2}{c|}{Simulation} \\
  $f_0$ (GHz)			& $\beta$		& $Q_0$		& $f_0$ (GHz)	& Q (Perturbation)		\\
  \hline
  \multicolumn{3}{|c|}{} & \multicolumn{2}{c|}{} \\
  0.384					& 0.096			& 114			& 0.379				& 482			\\
  0.772 		 		& 0.008			& 178			& 0.736				& 433			\\
  1.266			 		& 0.097			& 179			& 1.1310			& 525			\\
  1.366 		 		& 0.090			& 228			& 1.341				& 899			\\
  1.465      		   	& 0.037			& 1307  		& 1.487				& 1253			\\
  1.689      		   	& 0.055			& 216  			& 1.539				& 600			\\
  1.717      		   	& 0.039			& 209  			& 1.889				& 647			\\
  1.834      		   	& 0.067			& 246  			& 1.892				& 4708			\\
  1.986      		   	& 0.034			& 266	  		& 1.914				& 644			\\
  \multicolumn{3}{|c|}{} & \multicolumn{2}{c|}{} \\
  \hline  
\end{tabular}
\caption{Resonances measured using a probe for the cylindrical RP at 40 mm from the beam, without ferrite.}
\label{TableCyl40WOFerrite}
\end{table}

\begin{table}[htb!]
\centering
\begin{tabular}{ | c c c | c c | }
  \hline        
  \multicolumn{3}{|c|}{Probe measurement} & \multicolumn{2}{c|}{Simulation} \\
  $f_0$ (GHz)			& $\beta$		& $Q_0$		& $f_0$ (GHz)	& Q (Perturbation)	\\
  \hline
  \multicolumn{3}{|c|}{} & \multicolumn{2}{c|}{} \\
   				 			& 				& 			& 0.681				& 7.3		\\       
                 			& 				& 			& 0.902				& 8.9		\\
                 			& 				& 			& 1.162				& 15.0		\\
   		 		  			&  				&  			& 1.316				& 17.0		\\
  1.443					& 0.14			& 95.5		& 1.461				& 410.9		\\
                 			& 				& 			& 1.630				& 24.7		\\
  1.848					& 0.35			& 60		& 1.899				& 60.0		\\
   				 			& 				& 			& 1.98				& 0.72		\\
  \multicolumn{3}{|c|}{} & \multicolumn{2}{c|}{} \\
  \hline  
\end{tabular}
\caption{Resonances measured using a probe for the cylindrical RP at 10 mm from the beam, with ferrite.}
\label{TableCyl10WFerrite}
\end{table}

\clearpage
\section{Shielded RP}
\label{Shielded}
\begin{figure}[htb!]
\centering
\includegraphics[width=0.45\textwidth]{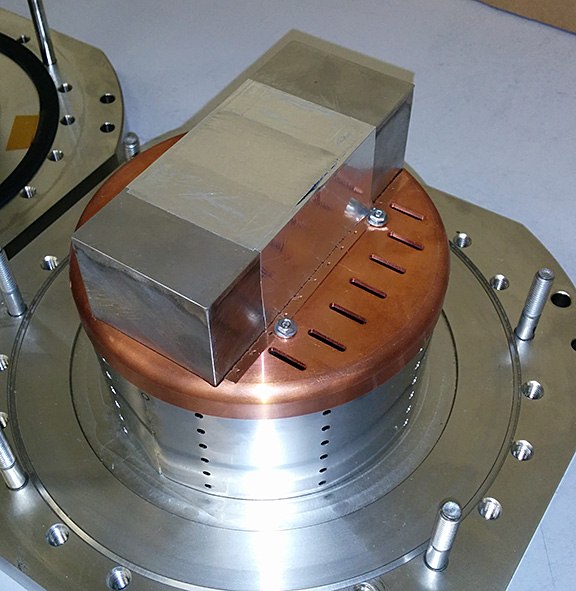}\quad
\includegraphics[width=0.45\textwidth]{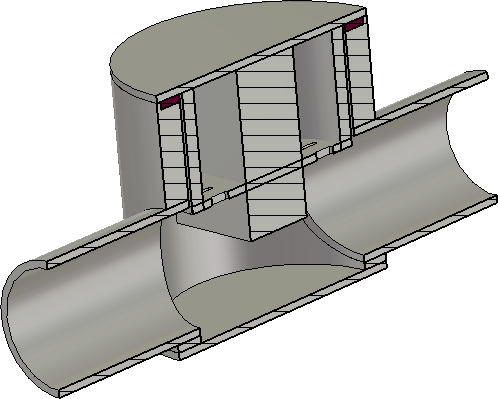}
\caption{The prototype used for the measurements and the model used for the simulations of the shielded RP. The copper shield is visible on the left picture and the ferrite ring is visible on the right picture.}
\label{ShieldedRP}
\end{figure}

Some of the old design RP will be inserted close to the LHC beam during operation at high intensity. A solution was therefore investigated in order to optimize the RP impedance without doing a completely new design.
The solution adopted consists in adding a ferrite ring and a copper shield to the existing RP, inside the primary vacuum of the LHC.
The shield is in vacuum and does not need to sustain high pressures, due to the presence of some holes.
The prototype used for the RF tests is shown in fig.~\ref{ShieldedRP}. It was not possible, for technical reasons, to test the shielded RP without the ferrite ring as in the previous case of Cylindrical RP.

\subsection{Single wire test}
As for the Cylindrical RP, the single wire tests were done with the RP in garage position (40 mm) and at 10 mm from the centre of the beam pipe.

\begin{figure}[htb!]
\centering
\begin{overpic}[width=0.9\textwidth]{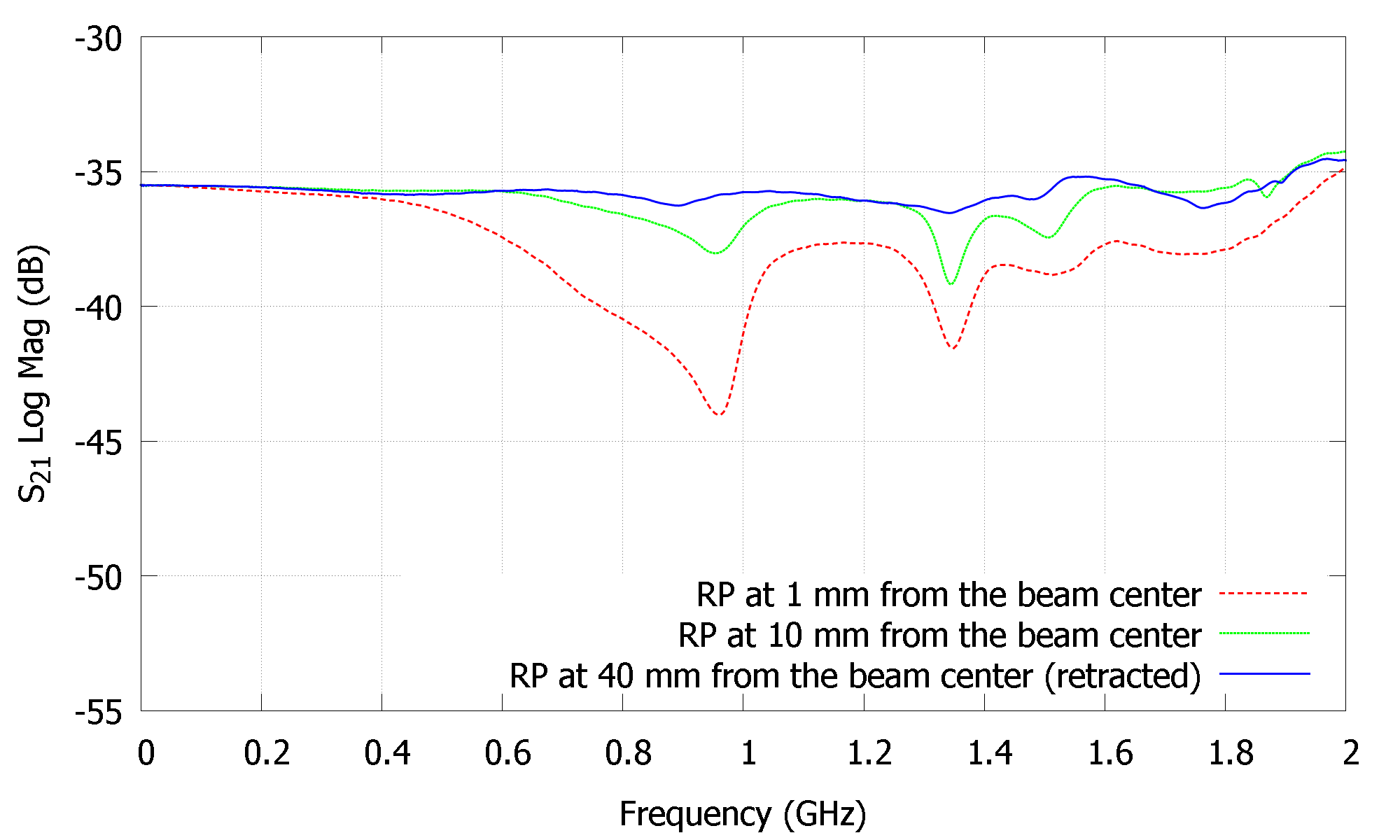}
\put (12,60) {\footnotesize Single wire measurements}
\end{overpic}
\caption{Measured transmission parameter $S_{21}$ for the wire test on Shielded RP with ferrite at 10 mm and 40 mm from the centre of the beam pipe.}
\label{ShieldedWFerrite}
\end{figure}

\begin{figure}[htb!]
\centering
\begin{overpic}[width=0.9\textwidth]{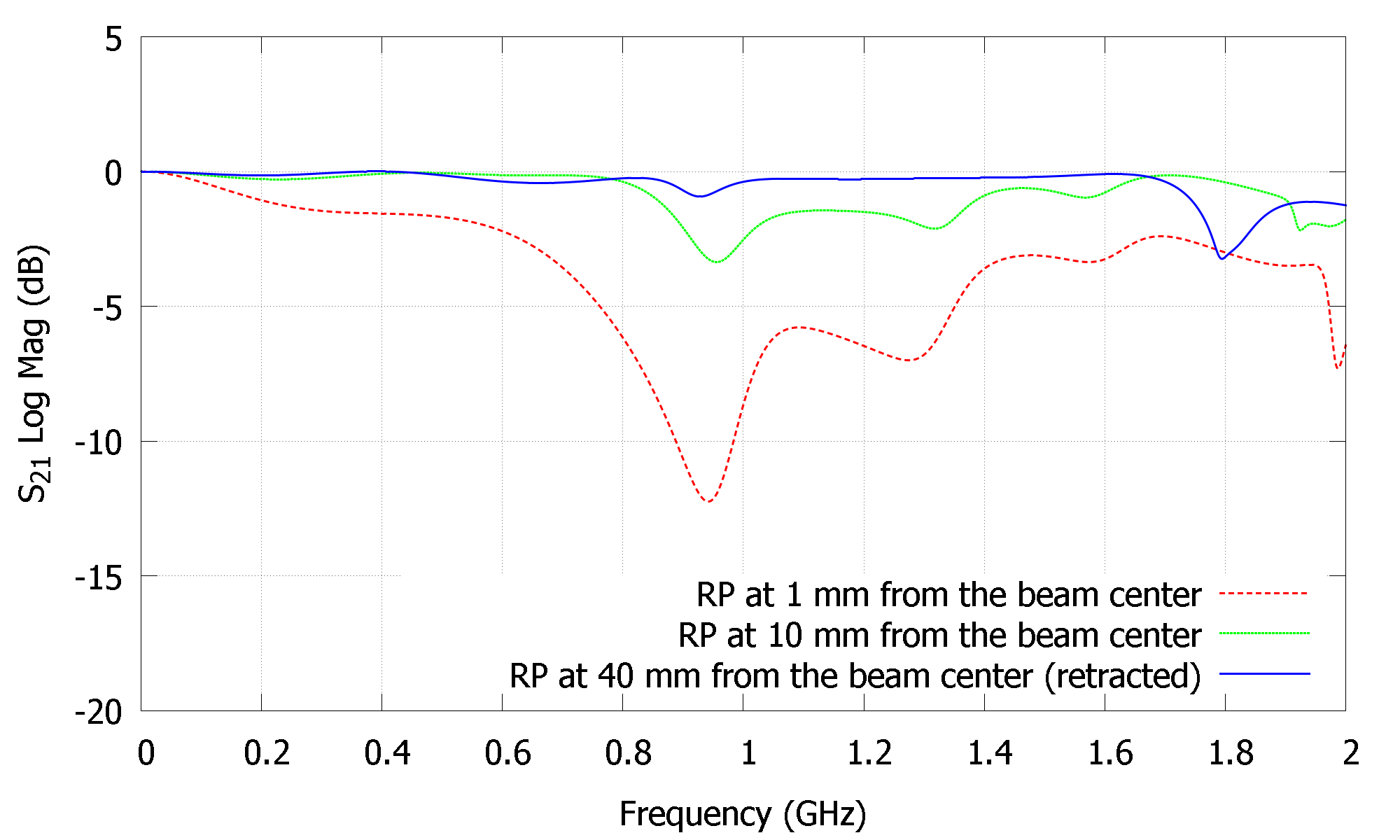}
\put (12,60) {\footnotesize Single wire simulations}
\end{overpic}
\caption{Simulated transmission parameter $S_{21}$ for the wire test on Shielded RP with ferrite at 10 mm and 40 mm from the centre of the beam pipe.}
\label{ShieldedWFerrite_SIM}
\end{figure}

Figures~\ref{ShieldedWFerrite} and~\ref{ShieldedWFerrite_SIM} show the measured and simulated $S_{21}$. The agreement between measurements and simulations is good both at 10 mm and 40 mm.

\clearpage
\subsection{Double wire test}
\begin{figure}[htb!]
\centering
\begin{overpic}[width=0.9\textwidth]{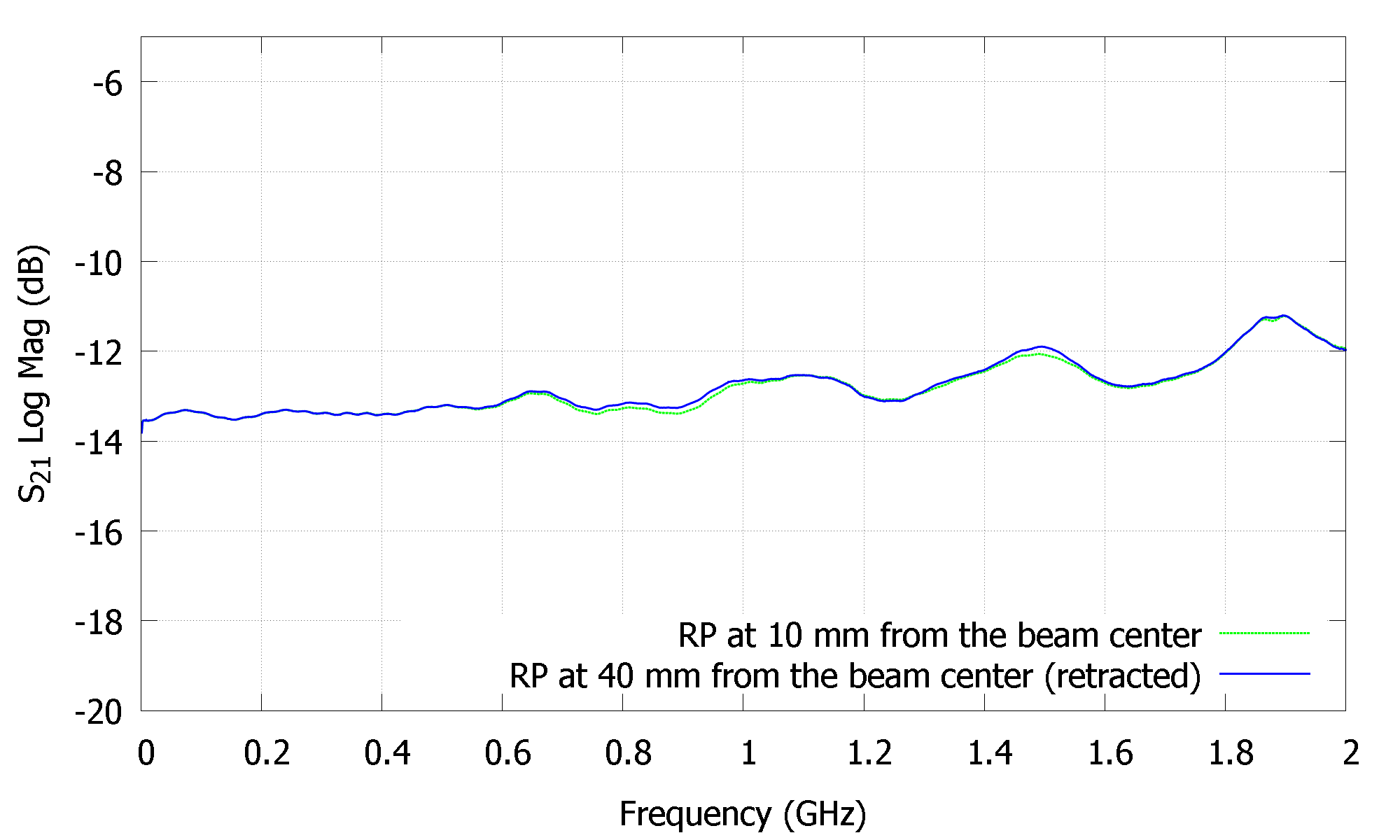}
\put (12,60) {\footnotesize Double wire horizontal, with ferrite}
\end{overpic}
\caption{Measured transmission parameter $S_{21}$ for the test with two wires horizontal with the Cylindrical RP at 10 mm and 40 mm from the centre of the beam pipe.}
\label{BoxShieldedWFerrite_HOR}
\end{figure}

\begin{figure}[htb!]
\centering
\begin{overpic}[width=0.9\textwidth]{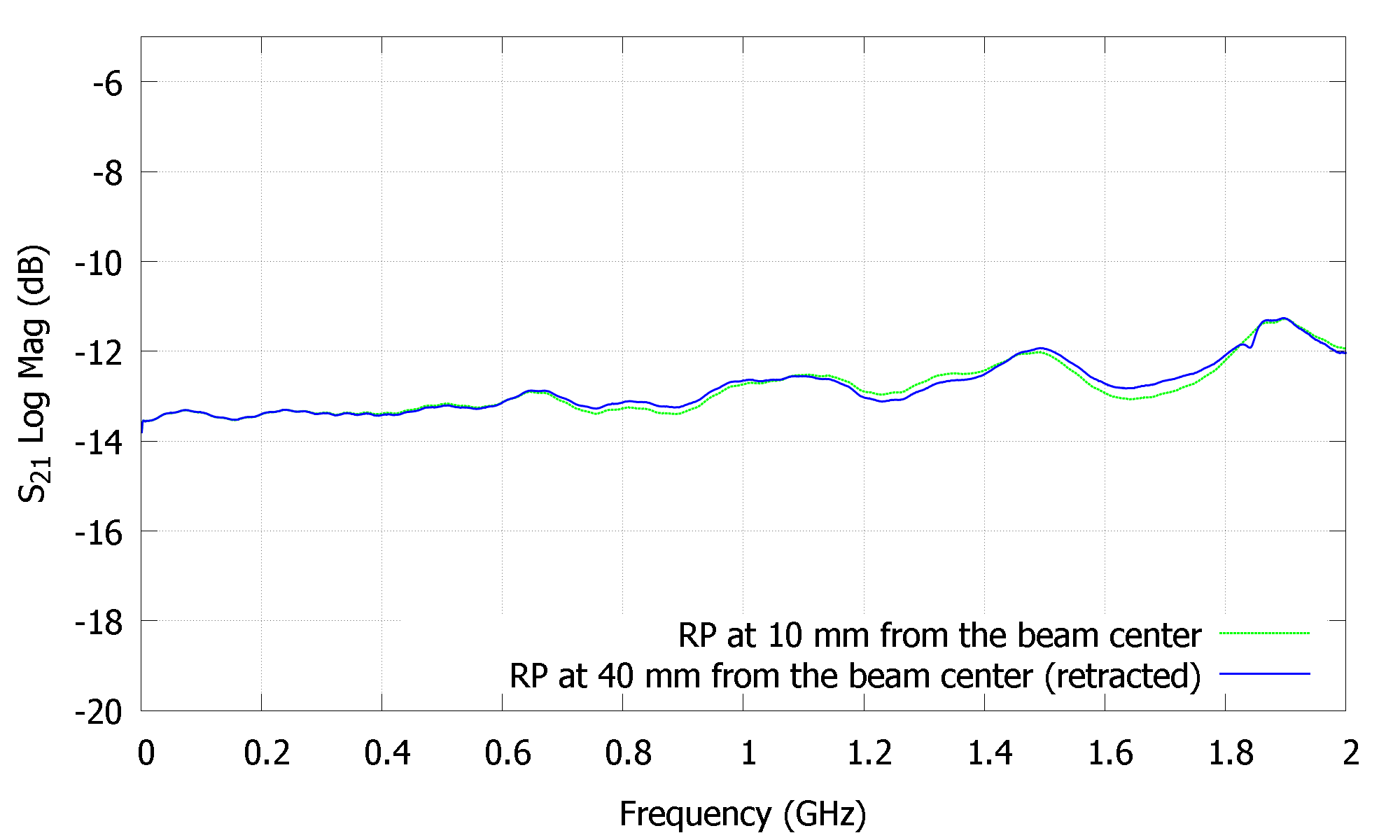}
\put (12,60) {\footnotesize Double wire vertical, with ferrite}
\end{overpic}
\caption{Measured transmission parameter $S_{21}$ for the test with two wires vertical with the Cylindrical RP at 10 mm and 40 mm from the centre of the beam pipe.}
\label{BoxShieldedWFerrite_VER}
\end{figure}

With analogy to the Cylindrical RP, double wire tests were performed. Figures~\ref{BoxShieldedWFerrite_HOR} and~\ref{BoxShieldedWFerrite_VER} show the measured $S_{21}$ on the shielded RP with ferrite. As for the Cylindrical RP, no resonance is visible below 1.8~GHz in both the horizontal and vertical plane.

\clearpage
\subsection{Probe test}
The probe test was done with the Shielded RP with the ferrite ring. As for the Cylindrical RP, in garage position it was not possible to detect any resonance. We summarize therefore the resonances detected at 10mm in table~\ref{TableShielded10WFerrite}.
The agreement between measurements and simulations is good: all the detected resonances have been foreseen by the simulations; some of them correspond to Q values that are too small to be measured with enough precision.

\begin{table}[htb!]
\centering
\begin{tabular}{ | c c c c | c c | }
  \hline        
  \multicolumn{4}{|c|}{Probe measurement} & \multicolumn{2}{c|}{Simulation (CST Eigenmode)} \\
  $f_0$ (GHz)	& $Q_L$		& $\beta$		& $Q_0$		& $f_0$ (GHz)	& $Q_0$ (Perturbation)	\\
  \hline
  \multicolumn{4}{|c|}{} & \multicolumn{2}{c|}{} \\
   				& 			& 				& 			& 0.63				& 7	\\
  0.983 		& 30		& 0.83			& 55		& 0.99				& 17	\\
  				& 			& 				& 			& 1.22				& 6	\\
  1.350			& 40		& 0.12			& 45		& 1.372				& 19	\\
   				& 			& 				& 			& 1.374				& 9
\\
  1.540 		& 21		& 0.51			& 33		& 1.63				& 12		\\
  1.865			& 40  		& 0.20  		& 48  		& 1.83				& 69
\\
  \multicolumn{4}{|c|}{} & \multicolumn{2}{c|}{} \\
  \hline  
\end{tabular}

\caption{Resonances measured and simulated using a probe for the shielded RP at 10 mm from the beam, with ferrite.}
\label{TableShielded10WFerrite}
\end{table}

\section{Conclusion}
The TOTEM Collaboration has designed, realized and tested a new Cylindrical Roman Pot and a new RF shield has been proposed and realized to improve the RF behavior of the old RP model without requiring a complete redesign.
The new design was simulated using CST Particle Studio and RF bench measurements were performed on the prototypes. The probe measurements and the single wire tests showed a good agreement with the simulations for both the Cylindrical RP and the Shielded RP. The two wires test confirmed the absence of relevant transverse resonances. 
The effectiveness of the ferrite in damping the resonances below 1.5 GHz has been demonstrated both in simulations and in measurements ensuring the equipment compatibility with the LHC requirements for safe operation.

\section{Acknowledgments}
We would like to express our gratitude to J.~Baechler, D.~Druzhkin, M.~Bozzo and M.~Deile, from the TOTEM collaboration, for their help and support while writing this note and E.~M\'{e}tral and H.~Burkhardt, from the BE-ABP group, for their supervision and for their interesting and masterful advice.

\bibliographystyle{unsrt}
\bibliography{biblio}

\end{document}